\documentclass[10pt]{article}

\usepackage{a4,amssymb,cite}

\newcommand{\rme}{{\rm e}}
\newcommand{\rmd}{{\rm d}}

\begin{document}

\begin{center}
  
  {\LARGE\bf Five-Dimensional Unification of\\[0.5cm] the Cosmological
    Constant and the Photon Mass}
  
  \vspace{0.7cm}
  
  {\large\sc Christopher Kohler}\footnote{E-mail: {\tt
      kohler@itap.physik.uni-stuttgart.de}}
  
  \vspace{0.5cm}

  {\small\it Institut f\"ur Theoretische und Angewandte Physik,}\\
  {\small\it Universit\"at Stuttgart,} {\small\it D-70550 Stuttgart, Germany}

\end{center}

\begin{abstract}
  Using a non-Riemannian geometry that is adapted to the 4+1 decomposition of
  space-time in Kaluza-Klein theory, the translational part of the connection
  form is related to the electromagnetic vector potential and a Stueckelberg
  scalar.  The consideration of a five-dimensional gravitational action
  functional that shares the symmetries of the chosen geometry leads to a
  unification of the four dimensional cosmological term and a mass term for
  the vector potential.
\end{abstract}


\section{Introduction}

Recently, an alternative formulation of Kaluza-Klein
theory~\cite{kaluza,klein} has been presented which is based on a differential
geometry that is adapted to the decomposition of the five-dimensional (5d)
Kaluza-Klein space-time into 4d space-time and the internal $S^{1}$
manifold~\cite{kohler1}. Physically, this approach leads to the same results
as conventional Kaluza-Klein theory except that it allows for the
Einstein-Cartan theory of 4d space-time.

In this article, the alternative formulation of Kaluza-Klein theory is
investigated further. In particular, the translational connection on the 5d
manifold is considered and it is shown that the part of it belonging to the
fifth dimension is related to the electromagnetic vector potential and a
Stueckelberg scalar. As a consequence, it will be seen that the $U(1)$ gauge
symmetry of Maxwell theory can be interpreted as a translational symmetry of
the internal space.

The use of the translational connection leads to a generalization of the
alternative Kaluza-Klein theory in that a mass term for the electromagnetic
field can be considered. The introduction of a 5d cosmological term that has
the same symmetries as the geometry used leads, after dimensional reduction,
to the Einstein-Cartan theory with cosmological constant coupled to massive
Maxwell theory in the Stueckelberg formulation thus unifying the cosmological
constant and the photon mass.

This article is organized as follows. In section 2, the alternative
formulation of Kaluza-Klein theory is described in the language of fibre
bundles. In section 3, the translational connection is introduced and its
relationship with the electromagnetic vector potential is revealed. The 5d
cosmological term and its dimensional reduction are considered in section 4.
Section 5 contains some conclusions.

In this article, the following conventions are employed. Uppercase indices
with values $0,1,2,3,5$ refer to 5d space-time where $A,B,\ldots$ are used for
internal indices and $M,N,\ldots$ for coordinate indices. Lowercase indices
with values $0,1,2,3$ are used for 4d space-time with $a,b,\ldots$ denoting
internal indices and $\mu,\nu,\ldots$ being coordinate indices. Explicit
indices in parentheses will always be internal indices. The metric signature
is $(-+++(+))$. Throughout this article, geometrized units are used where
$16\pi G=1=c$.

\section{Five-Dimensional Unification of Einstein-Car\-tan Theory and 
  Max\-well Theory}

Kinematically, Kaluza-Klein theory is based on two assumptions: First, the 5d
space-time manifold $M^{5}$ is decomposed into the 4d space-time manifold
$M^{4}$ and a 1-sphere $S^{1}$ representing the internal space, that is, $
M^{5} = M^{4} \times S^{1} $. Secondly, all geometrical fields defined on
$M^{5}$ have a dependence only on the points of $M^{4}$. (We disregard massive
excitations of the geometry.)  Assuming a 5d local Lorentz gauge symmetry on
$M^{5}$, these two restrictions can be expressed as symmetry breakings: The
4+1 decomposition of $M^5$ is described by the symmetry breaking $SO(4,1)
\rightarrow SO(3,1)$ on $M^5$ where $SO(3,1)$ represents the local Lorentz
symmetry on $M^{4}$. The fact that the geometrical fields only depend on 4d
spacetime corresponds to the symmetry breaking $SO(3,1) \rightarrow 1$ on each
single internal manifold $S^1$.

Starting from a general Riemann-Cartan geometry on $M^5$, the consequent
application of standard theorems for the reduction of linear
connections~\cite{kobayashi} on the symmetry breakings leads to a unique
geometry on $M^5$. This will be explained in this section. The reduction is
described in some detail in order to be able to generalize it to affine
connections in the following section.

A basis of tangent vectors $e_A = e_{A}^{M}\partial_{M}$ at a point of $M^5$
forms a linear frame. The set of all linear frames at all points of $M^5$
forms the bundle of linear frames $L(M^5)$. The restriction to orthonormal
frames, which here means frames transforming under $SO(4,1)$, amounts to the
choice of a 5d metric tensor $\gamma_{MN}$ with respect to which the frames
are orthonormal. The corresponding fibre bundle is the bundle of orthonormal
frames $O(M^{5})$. The generators of the Lie algebra $so(4,1)$ of $SO(4,1)$
will be denoted by $J^{AB}$ and satisfy the Lie algebra
\begin{equation} \label{2.1} 
  [ J^{AB}, J^{CD} ] = 2\eta^{C[A} J^{B]D} - 2\eta^{D[A} J^{B]C}
\end{equation}
where $\eta^{AB}$ is the 5d Minkowski metric and square brackets denote
antisymmetrization.  Each element $A = \frac{1}{2} A_{AB} J^{AB}$ of $so(4,1)$
induces a so called fundamental vector field $A^{\ast}$ on $O(M^{5})$ which is
given by
\begin{equation} \label{2.2} 
  A^{\ast} = A_{AB} e^{AM} \frac{\partial}{\partial e^M_B}
\end{equation}
with $e^{AM}\equiv\eta^{AB}e^{M}_{B}$.  The integral curves of $A^{\ast}$ are
1-parameter groups of $SO(4,1)$ transformations on the fibres of $O(M^{5})$
generated by $A$.  A connection form $\omega$ on $O(M^5)$ is an $so(4,1)$
valued 1-form on $O(M^5)$ that satisfies $\omega(A^{\ast}) = A$ and transforms
under global $SO(4,1)$ transformations of the frames according to the adjoint
representation.  From equation (\ref{2.2}) then follows that $\omega$ is of
the form
\begin{equation} \label{2.3} 
  \omega = \frac{1}{2} \left( e_{MA} \rmd e^{M}_{B} + 
    e_{MA} e^{N}_{B} \Gamma^{M}_{NP} \rmd x^P \right)
  J^{AB}
\end{equation}  
where the functions $\Gamma^M_{NP}$ are the connection coefficients and
$e_{AM} \equiv \eta_{AB}e^{B}_{M}$ with the coframe $e^{A}$ which is
determined by the duality relation $e^{A}(e_{B}) = \delta^{A}_{B}$.  Under a
local $SO(4,1)$ transformtion $e'{\!}_{A} = e_{B}\Omega^{B}{}_{A}$ of the
frames, the connection form $\omega$ (or, more precisely, its pullback to
$M^{5}$) transforms as
\begin{equation}\label{2.3a}
  \omega^{A}{}_{B} \rightarrow \Omega^{-1}{}^{A}{}_{C} 
  \omega^{C}{}_{D} \Omega^{D}{}_{B} +
  \Omega^{-1}{}^{A}{}_{C} \rmd \Omega^{C}{}_{B} .
\end{equation}

The 4+1 decomposition of $M^5$ allows the use of special frames $e_A$ whose
basis vectors $e_{(5)}$ point in the fifth dimension, that is, $e_{(5)}$ is a
multiple of $\partial_{5}$. These frames form a subbundle $Q(M^5)$ of $O(M^5)$
with structure group $SO(3,1)$. It is defined by the tensor
\begin{equation} \label{2.4} 
  q^{M}_{N} = e^M_{a} e^{a}_{N} 
\end{equation}
which projects 5d vectors onto the 4d spacetime manifold $M^{4}$ and induces a
metric on $M^4$. With respect to $O(M^{5})$, this tensor can be regarded as a
section of the associated fibre bundle with standard fibre $SO(4,1)/SO(3,1)$.
The connection (\ref{2.3}) is reducible to a connection $\omega '$ on $Q(M^5)$
if and only if $q^{M}_{N}$ is parallel with respect to $\omega$, from which
follows
\begin{equation} \label{2.5}
  \omega_{a(5)M} = - e^N_a e^P_{(5)} D_M q_{NP} = 0 
\end{equation}
where $D_M$ denotes the covariant derivative corresponding to $\omega$ and
$q_{MN} \equiv \gamma_{MP}q^{P}_{N}$.

The requirement to be able to perform a dimensional reduction means that
geometrical quantities do not depend on the coordinate $x^{5}$ of $S^{1}$.
This implies that a preferred basis $e_{a}$ can be chosen where $e_{a}$ is
fixed along each internal space belonging to a point of $M^{4}$. Since each
internal space $S^{1}$ at a point $x^{\mu}$ defines a subbundle
$Q(S^{1},x^{\mu})$ of $Q(M^{5})$ by restricting $Q(M^{5})$ to the fibres over
the $S^{1}$ manifold at the point $x^{\mu}$ of $M^{4}$, the preferred frames
form a subbundle $S(S^{1},x^{\mu})$ of $Q(S^{1},x^{\mu})$ with structure
group $1$.  The preferred frames $e_{a}(x^{5},x^{\mu})$ can be considered as a
section of the associated fibre bundle over $S^{1}$ with standard fibre
$SO(3,1)$.  The $SO(3,1)$ connection $\omega '$ induces a connection on each
subbundle $Q(S^{1},x^{\mu})$ of $Q(M^{4})$. These connections are reducible to
connections on $S(S^{1},x^{\mu})$ if and only if $e_{a}(x^{5},x^{\mu})$ is
parallel with respect to the induced connection, which means
\begin{equation} \label{2.6}
  \omega_{ab5} = e^{M}_{a} D_{5} e_{bM} = 0
\end{equation}

The conditions (\ref{2.5}) and (\ref{2.6}) define a connection on $M^{5}$ that
is adapted to the symmetries of the Kaluza-Klein reduction. It was previously
called a semi-teleparallel connection~\cite{kohler1} since it parallelizes the
internal space while leaving the Riemann-Cartan connection on $M^{4}$
unrestricted.

We will base a five dimensional unification on this connection instead as on
the Riemannian connection of conventional Kaluza-Klein theory. However,
formally we will follow the same procedure in chosing an action functional as
in Kaluza-Klein theory.  To start with, we observe that similar to the unique
decomposition
\begin{equation} \label{2.7}
  \omega^A{}_{BM} = \;\stackrel{\circ}{\omega}{\!\!}^A{}_{BM} - K^A{}_{BM}
\end{equation}
of a Lorentz connection $\omega^{A}{}_{BM}$ into the Levi-Civita connection
$\stackrel{\circ}{\omega}{\!\,\!\!\!}^A{}_{BM}$ and the contortion tensor
$K^A{}_{BM}$, we have a unique decomposition~\cite{kohler2}
\begin{equation} \label{2.8}
  \omega^A{}_{BM} = \tilde{\omega}^A{}_{BM} + S^A{}_{BM}
\end{equation}
of a Lorentz connection $\omega^{A}{}_{BM}$ into a semi-teleparallel
connection $\tilde{\omega}^A{}_{BM}$ and a tensor field $S^A{}_{BM}$ that is
restricted by the condition that its projection onto $M^{4}$ vanishes,
\begin{equation} \label{2.8.1}
  q^{M}_{Q} q^{N}_{R} q^{P}_{S} S_{MNP} = 0
\end{equation}
where $S_{MNP} \equiv e^{A}_{M} e^{B}_{N}S_{ABP}$.

In order to obtain a unique action for the semi-teleparallel connection, we
start with the Einstein-Cartan action on $M^5$,
\begin{equation}\label{2.9.1}
  S[e^{A}, \omega_{AB}] = \int_{M^5} {\rm d}^{5}x\;
  \sqrt{-\gamma}\; 
  R(e^{A}, \omega_{AB}) .
\end{equation}
Here, $\gamma$ is the determinant of the 5d metric $\gamma_{MN}$ and $R$ is
the 5d scalar curvature. We then insert the decomposition (\ref{2.8}) for the
Lorentz connection and determine the stationary point with respect to the
tensor field $S^A{}_{BM}$. As a result, we obtain a functional of the
semi-teleparallel connection $\tilde{\omega}$. In the case of the
decomposition (\ref{2.7}), this procedure leads to the Einstein-Hilbert action
which is chosen as the action functional in conventional Kaluza-Klein theory.
Thus, we expect by using the decomposition (\ref{2.8}) to arrive at a sensible
action functional for a 5d unification.

The determination of the stationary point of the action functional is best
done using the preferred frames $e_{A}$ that are adapted to the Kaluza-Klein
geometry since in this basis we have $\omega_{a(5)M}=0=\omega_{ab5}$ and the
condition (\ref{2.8.1}) means $S_{abc}=0$. Inserting the decomposition
(\ref{2.8}) into the Einstein-Cartan action and varying with respect to
$S_{ab(5)}$ and $S_{(5)ab}$, we obtain
\begin{eqnarray}\label{2.10}
  S_{ab(5)} & = & \tilde{K}_{ab(5)} \\ \label{2.10.1}
  S_{(5)ab} & = & \tilde{K}_{(5)ab}
\end{eqnarray}
where $\tilde{K}_{ABC}$ is the contortion tensor of the semi-teleparallel
connection $\tilde{\omega}_{ABC}$.  The Einstein-Cartan action contains the
components $S_{(5)a(5)}$ only within the linear term $S_{(5)a(5)}
\tilde{K}^{ab}{}_{b}$.  Hence, the action possesses a stationary point only if
the 4d torsion is tracefree, $\tilde{K}^{ab}{}_{b}=0$, which is the case in
the absence of spinning matter.  Reinserting (\ref{2.10}) and (\ref{2.10.1})
into the Einstein-Cartan action, we obtain the action
\begin{equation} \label{2.11}
  S[e^{A}, \tilde{\omega}_{AB}] = 
  \int_{M^{5}} \rmd^{5}x \sqrt{-\gamma} \left(
    {}^{4}\!R -\frac{1}{4} T^{(5)ab} T_{(5)ab} \right) .
\end{equation}
Here, ${}^{4}\!R$ is the 4d scalar curvature and $T^{A}{}_{BC}$ is the 5d
torsion tensor of $\tilde{\omega}$.  We choose the action functional
(\ref{2.11}) as the action for the alternative five dimensional unification.

In order to make contact with electromagnetism, we parameterize the 5d coframe
$e^{A}$.  The fifth component of $e_{A}$ is chosen to be
\begin{equation} \label{2.12}
  e_{(5)} = \rme^{-\phi(x^{\mu})} \partial_{\theta}
\end{equation}
where $\theta \equiv x^{5}$ is the coordinate along the internal $S^{1}$ and
$\phi(x^{\mu})$ is a scalar field as usual in the Kaluza-Klein theory. From
the duality relation, the fifth component of the coframe then has the form
\begin{equation}\label{2.13}
  e^{(5)} = \rme^{\phi} \rmd\theta + C_{\mu} \rmd x^{\mu}
\end{equation} 
where $C_{\mu}$ is a 4d vector field. The remaining components of the basis
and cobasis follow from equations (\ref{2.12}) and (\ref{2.13}) using the
duality relation:
\begin{eqnarray}\label{2.14}
  e_{a} & = & e_{a}^{\mu} \partial_{\mu} -e_{a}^{\mu} C_{\mu} 
  \rme^{-\phi}\partial_{\theta}  \\ \label{2.14.1}
  e^{a} & = & e^{a}_{\mu} \rmd x^{\mu} .
\end{eqnarray}
As in conventional Kaluza-Klein theory, the vector field $C_{\mu}$ shall be
related to the electromagnetic vector potential $A_{\mu}$. To find this
relationship, we note that in Kaluza-Klein theory gauge transformations of
$A_{\mu}$ are attributed to coordinate transformations along the internal
space, that is, $\theta \rightarrow \theta +\lambda(x^{\mu})$. Inserting this
transformation into equation (\ref{2.13}) leads to
\begin{equation} \label{2.15}
  e^{(5)} = \rme^{\phi}\rmd\theta 
  + \left( \rme^{\phi} \partial_{\mu} \lambda +
    C_{\mu}\right) \rmd x^{\mu} .
\end{equation}
From this follows that we have to choose 
\begin{equation}\label{2.15.1}
  C_{\mu} = \rme^{\phi} A_{\mu}
\end{equation}
in order to obtain the correct gauge transformation
$A_{\mu}\rightarrow A_{\mu} + \partial_{\mu}\lambda$.

Using the chosen parameterization of the basis as the preferred frames for the
semi-teleparallel connection, the fifth component of the torsion tensor is
given by
\begin{equation}\label{2.16}
 T^{(5)}{}_{ab} = \rme^{\phi} e^{\mu}_{a} e^{\nu}_{b} F_{\mu\nu}
\end{equation}
where $F_{\mu\nu} = \partial_{\mu} A_{\nu} - \partial_{\nu} A_{\mu}$
represents the electromagnetic field strength. Then, the action functional
(\ref{2.11}) reads after dimensional reduction
\begin{equation}\label{2.17}
  S[e^{a}, \omega_{ab}, A_{\mu}, \phi] = \int_{M_{4}} \rmd^{4}x \sqrt{-g} 
  \rme^{\phi}\left( {}^{4}\!R - \frac{1}{4} \rme^{2\phi} F_{\mu\nu}
  F^{\mu\nu} \right)
\end{equation}
where $g$ is the determinant of the 4d metric $g_{\mu\nu} =
e^{a}_{\mu}e^{b}_{\nu}\eta_{ab}$. Equation (\ref{2.17}) coincides with the
action of the Einstein-Cartan-Maxwell theory coupled to a scalar field.

\section{Translational Connection} \label{trans}

We next generalize the linear connection considered in the previous section to
an affine connection, that is, we additionally introduce a translational
connection on $M^{5}$~\cite{kobayashi}. Applying symmetry breakings analogous
to the ones used to define the semi-teleparallel connection, we will see that
the translational connection has a natural relation with the electromagnetic
vector potential.

The bundle of linear frames $L(M^{5})$ can be extended to the bundle of affine
frames $A(M^{5})$ by generalizing linear frames $e^{M}_{A}$ to affine frames
$(e^{M}_{A},q^{M})$ which include affine vectors $q^{M}=e^{M}_{A}q^{A}$ at
each point of $M^{5}$. In the case of the bundle of orthonormal frames
$O(M^{5})$, the corresponding bundle of affine frames $P(M^{5})$ consists of
frames that transform under the 5d Poincar\'{e} group $P(5) = SO(4,1) \ltimes
T^{5}$. $P(5)$ is generated by the generators $(J^{AB}, P_{A})$ which have,
besides equation (\ref{2.1}), the Lie brackets
\begin{eqnarray}\label{3.1}
  {}[ J^{AB}, P_{C} ]   & = & 2 \delta^{[B}_{C} \eta^{A]D} P_{D}  \\
  {}[P_{A}, P_{B} ]     & = & 0 .
\end{eqnarray}
Under a Poincar\'{e} transformation, an affine frame transforms according to
\begin{equation}\label{3.1a}
  \left( e_{A}^{M},q^{A} \right) \rightarrow \left( \Omega^{B}{}_{A} e^{M}_{B},
    \Omega^{-1}{}^{A}{}_{B} ( q^{B} - t^{B}) \right)
\end{equation}
where $t^{A}$ is a translation.  The fundamental vector field $A^{\ast}$ on
$P(M^{5})$ corresponding to the Lie algebra element $A=\frac{1}{2} A_{AB}
J^{AB} + A^{A} P_{A}$ is
\begin{equation}\label{3.2}
  A^{\ast} = A_{AB} e^{AM}\frac{\partial}{\partial e^{M}_{B}} 
  - A^{A} e^{M}_{A} \frac{\partial}{\partial q^{M}} .
\end{equation}
From this follows the connection form
\begin{equation}\label{3.3}
  \hat{\omega} = \omega + \varphi
\end{equation}
with the translational connection form
\begin{equation}\label{3.4}
  \varphi = e^{A}_{M} \left( - \rmd q^{M} - 
    \Gamma^{M}_{NP} q^{N} \rmd x^{P} +
    \Upsilon^{M}_{N}\rmd x^{N} \right) P_{A}
\end{equation}
where $\omega$ and $\Gamma^{M}_{NP}$ are the $SO(4,1)$ connection form and
connection coefficients, respectively, defined in equation (\ref{2.3}), and
$\Upsilon^{M}_{N}$ are the connection coefficients of the translational
connection.  Under a local Poincar\'{e} transformation, $\varphi^{A}$
transforms as
\begin{equation}\label{3.4a}
  \varphi^{A} \rightarrow \Omega^{-1}{}^{A}{}_{B} 
  \left( \varphi^{B} + {\rm D}t^{B}\right) 
\end{equation}
where ${\rm D}t^{A} \equiv \rmd t^{A} + \omega^{A}{}_{B} t^{B} $.  Since
$\varphi^{A}$ transforms under local Lorentz transformations as a tensor, we
will set $\Upsilon^{M}_{N} = \delta^{M}_{N}$ in the following thus relating
$\varphi^{A}$ to the cobasis $e^{A}$. The translational connection form can
then be written as
\begin{equation}\label{3.5}
  \varphi^{A} = - {\rm D}q^{A} + e^{A} .
\end{equation}
Equation (\ref{3.5}) was first introduced in reference~\cite{trautman}. We
will follow reference~\cite{grignani} and refer to the bundle coordinates
$q^{A}$ as Poincar\'{e} coordinates.

We next apply the symmetry reductions described in the previous section to the
translational connection. The first symmetry breaking $SO(4,1)\rightarrow
SO(3,1)$ generalizes to
\begin{equation}\label{3.6}
  P(5) \rightarrow SO(3,1)\ltimes T^{5} .
\end{equation}
The 5d translational symmetry is not broken since the 4+1 decomposition of
space-time can be realized by splitting the translations into 4d space-time
translations and internal ones. The reduced subbundle of $P(M^{5})$ that
corresponds to (\ref{3.6}) will be denoted by $R(M^{5})$. The translational
part of $R(M^{5})$ is not changed with respect to $P(M^{5})$. Accordingly, the
translational connection is not affected by the symmetry breaking (\ref{3.6}).
However, the second symmetry breaking, which is restricted to the fibres
$R(S^{1},x^{\mu})$ of the bundle $R(M^{5})$, generalizes to
\begin{equation}\label{3.7}
  SO(3,1) \ltimes T^{5} \rightarrow 1 \times T^{1}
\end{equation}
where $T^{1}$ corresponds to the internal $S^{1}$ translations which are
$U(1)$ transformations. The reason for (\ref{3.7}) is that 4d space-time
translations --- like 4d space-time Lorentz transformations --- must be
independent of the coordinates of the internal space in order to be able to
perform the dimensional reduction.

The induced connection form on $R(S^{1},x^{\mu})$ is
\begin{equation}\label{3.8}
  \hat{\omega} = \frac{1}{2} \omega_{ab5} \rmd\theta J^{ab} 
  + \varphi^{A}_{5} \rmd\theta P_{A} . 
\end{equation}
The symmetry breaking (\ref{3.7}) means that the connection (\ref{3.8}) is
reducible to the $T^{1}$ connection $\hat{\omega}=\varphi^{(5)}_{5} \rmd\theta
P_{5}$ on each fibre. Consequently, besides $\omega_{ab5}=0$, the components
$\varphi^{a}_{5}$ of the translational connection vanish.

Next we seek a physical interpretation of the semi-teleparallel translational
connection. In accordance with the Kaluza-Klein theory, we require that the
fields $e^{A}_{M}$ and $q^{A}$ only depend on the 4d space-time. (A dependence
of $q^{(5)}$ on $\theta$ can be removed by a $T^{1}$ gauge transformation.)
According to equation (\ref{3.5}), the 4d space-time part of the coframe is
given by
\begin{equation}\label{3.9}
  e^{a}_{\mu} = \partial_{\mu} q^{a} + \omega^{a}{}_{b\mu} q^{b} + 
  \varphi^{a}_{\mu} .
\end{equation}
This follows from the fact that $\omega^{a}{}_{(5)\mu}$ vanishes in a
semi-teleparallel geometry. Equation (\ref{3.9}) is the known relation between
the 4d cobasis $e^{a}_{\mu}$, the Riemann-Cartan connection
$\omega^{a}{}_{b\mu}$, the 4d Poincar\'{e} coordinates $q^{a}$ and the 4d
translational connection $\varphi^{a}_{\mu}$. Thus, equation (\ref{3.9}) can
be understood within a 4d Poincar\'{e} gauge theory~\cite{grignani}.

For the components $e^{a}_{5}$ of the coframe, we have from equation
(\ref{3.5})
\begin{equation}\label{3.10}
  e^{a}_{5} = \partial_{5} q^{a} + \omega^{a}{}_{B5} q^{B} 
  + \varphi^{a}_{5} = 0
\end{equation}
because $\omega^{a}{}_{B5}=0$ and $\varphi^{a}_{5}=0$ for a semi-teleparallel
geometry and because $q^{a}$ is independent of the coordinate $\theta$ of the
internal space. Equation (\ref{3.10}) is in accordance with our choice of the
adapted frame (\ref{2.14.1}).

The component $e^{(5)}_{5}$ is given by
\begin{equation}\label{3.11}
  e^{(5)}_{5} = \partial_{5} q^{(5)} + \omega^{(5)}{}_{a5} q^{a} +
  \varphi^{(5)}_{5} = \varphi^{(5)}_{5}
\end{equation}
since $q^{(5)}$ does not depend on $\theta$ and $\omega^{(5)}{}_{a5}=0$.
Comparing equation (\ref{3.11}) with equation (\ref{2.13}), we find
\begin{equation}\label{3.12}
  \varphi^{(5)}_{5} = \rme^{\phi} .
\end{equation}

Finally, for the components $e^{(5)}_{\mu}$ follows
\begin{equation}\label{3.13}
  e^{(5)}_{\mu} = \partial_{\mu} q^{(5)} + \omega^{(5)}{}_{a\mu} q^{a} +
  \varphi^{(5)}_{\mu} = \partial_{\mu} q^{(5)} + \varphi^{(5)}_{\mu} . 
\end{equation}
Comparison with equation (\ref{2.13}) yields
\begin{equation}\label{3.14}
  \varphi^{(5)}_{\mu} = C_{\mu} - \partial_{\mu}\sigma
\end{equation}
where we have set $\sigma\equiv q^{(5)}$. In order to find the physical
meaning of this equation, we consider a $T^{1}$ gauge transformation
$\sigma\rightarrow \sigma - \lambda(x^{\mu})$. From equations (\ref{3.4a}) and
(\ref{3.14}) follows
\begin{equation}\label{3.15}
  \varphi^{(5)}_{\mu} \rightarrow \varphi^{(5)}_{\mu} + \partial_{\mu}\lambda
\end{equation}
which is a $U(1)$ gauge transformation. Hence, we can identify the
translational connection $\varphi^{(5)}_{\mu}$ with the electromagnetic vector
potential $A_{\mu}$. Then, $\sigma$ has the meaning of a Stueckelberg
scalar~\cite{stueckelberg}.  Note that the identification
\begin{equation}\label{3.16}
  C_{\mu}=A_{\mu}+\partial_{\mu}\sigma
\end{equation}
differs from (\ref{2.15.1}) in that coordinate transformations of the internal
space no longer induce $U(1)$ transformations --- provided $\phi$ is not zero.
Instead, the $U(1)$ gauge transformations are translational gauge
transformations.

The construction of the action functional given in the previous section
remains unchanged by the introduction of the translational connection. This
follows from the fact that the coframe $e^{A}$ is not affected by the symmetry
reductions and there is no explicit dependence of the action functional on the
translational connection.

To summarize this section, it has been shown that the electromagnetic vector
potential $A_{\mu}$ can be interpreted as the fifth component
$\varphi^{(5)}_{\mu}$ of a translational connection where $U(1)$ gauge
transformations correspond to internal $S^{1}$ translations. Furthermore, the
fifth component $q^{(5)}$ of the Poincar\'{e} coordinates can be interpreted
as a Stueckelberg scalar $\sigma$.

It should be remarked that these identifications cannot be performed by
generalizing a 5d Riemann-Cartan geometry or a 5d Riemannian geometry to
include translations.

\section{Cosmological Term}

In this section, we consider the addition of a cosmological term to the action
(\ref{2.17}).  In Kaluza-Klein theory, the cosmological term is
\begin{equation}\label{4.1}
  \Lambda \int_{M^{5}} \rmd^{5} x \sqrt{-\gamma}
\end{equation}
where $\Lambda$ is the cosmological constant. Using differential forms,
(\ref{4.1}) may be rewritten as
\begin{equation}\label{4.2}
  \Lambda_{AB} \int_{M^{5}} e^{A} \wedge \ast e^{B} .
\end{equation}
Here, $e^{A}$ is a 5d orthonormal coframe, $\Lambda_{AB}$ is a constant
tensor, and $\ast$ is the duality operator defined by
\begin{equation}\label{4.3}
  \alpha \wedge \ast \beta = ( \alpha, \beta ) e
\end{equation}
where $\alpha$,$\beta$ are p-forms, $(\cdot , \cdot)$ denotes the scalar
product defined by the 5d metric induced by the coframe $e^{A}$, and $e$ is
the corresponding volume form. 5d Lorentz invariance requires the tensor
$\Lambda_{AB}$ to be of the form
\begin{equation}\label{4.4}
  \Lambda_{AB} = \frac{\Lambda}{5} \eta_{AB} .
\end{equation}
Indeed, using equation (\ref{4.3}) in (\ref{4.2}) yields (\ref{4.1}) where
$\Lambda_{AB}$ and $\Lambda$ are related by equation (\ref{4.4}).

The main difference between the approach to 5d unification presented in this
work and traditional Kaluza-Klein theory is the breaking of the 5d Lorentz
invariance to a 4d Lorentz invariance. Therefore, we have also to break down
the 5d Lorentz invariance of the cosmological term. This can be done by using
a different duality operator in (\ref{4.2}). We can define a new duality
operator $\sharp$ by
\begin{equation}\label{4.5}
   e^{A}\wedge \sharp e^{B} = e^{A}_{M}e^{B}_{N}\;
   {}^{\sharp}\gamma^{MN}\sqrt{-{}^{\sharp}\gamma}\;
   {\rm d}^{5}x
\end{equation}
where ${}^{\sharp}\gamma^{MN}$ is a 5d metric different from $\gamma^{MN}$.
Since we want to retain 4d Lorentz invariance, equation (\ref{4.5}) has to be
invariant under 4d Lorentz transformations which means that the 4d part of
${}^{\sharp}\gamma^{MN}$ is the one of $\gamma^{MN}$, that is,
${}^{\sharp}\gamma^{\mu\nu} = g^{\mu\nu}$. For ${}^{\sharp}\gamma^{MN}$ we use
the parameterization
\begin{equation}\label{4.6}
  {}^{\sharp}\gamma^{MN} =
   \left(
     \begin{array}{cc}
       g^{\mu\nu} & -B^{\nu} \\
       -B^{\mu}   & \rme^{-2\psi}+B^{\mu}B_{\mu}
     \end{array}
   \right)
\end{equation}
where $B^{\mu}$ is a 4d vector field and $\psi$ is a scalar field.

The restricted invariance under 4d Lorentz transformations requires
$\Lambda_{AB}$ to be of the form
\begin{equation}\label{4.7}
  \Lambda_{AB} = 
  \left(
    \begin{array}{cc}
      \frac{\Lambda}{4}\eta_{ab} & 0 \\
      0 & \Sigma
    \end{array}
  \right)
\end{equation}
where $\Sigma$ is a constant. Using this equation and the modified duality
operator $\sharp$, (\ref{4.2}) yields after dimensional reduction
\begin{eqnarray}\label{4.8}
  \lefteqn{ \int_{M^{4}} \rmd^{4} x \sqrt{-g} \rme^{\psi}
  \Big[ \Lambda_{ab}e^{a}_{\mu}e^{b}_{\nu}g^{\mu\nu}} \nonumber \\
  & + & \Sigma e^{(5)}_{\mu}e^{(5)}_{\nu}g^{\mu\nu} - 2\Sigma B^{\mu}
    e^{(5)}_{\mu}e^{(5)}_{5}
    + \Sigma e^{(5)}_{5}e^{(5)}_{5}\left( 
      \rme^{-2\psi}+B^{\mu}B_{\mu}\right) \Big] .
\end{eqnarray}
If we insert the $e^{(5)}$ given by equations (\ref{2.13}) and (\ref{3.16}),
and use that $g^{\mu\nu} = e_{a}^{\mu}e_{b}^{\nu}\eta^{ab}$, we obtain
\begin{eqnarray}\label{4.9}
  \lefteqn{ \int_{M^{4}} \rmd^{4} x \sqrt{-g} \rme^{\psi} 
  \Big[ \Lambda + \Sigma \rme^{2\phi - 2\psi}} \nonumber \\
   & + & \Sigma  
    \left(A_{\mu} + \partial_{\mu} \sigma- \rme^{\phi}B_{\mu} \right)
    \left( A_{\nu} + \partial_{\nu} \sigma 
      - \rme^{\phi}B_{\nu} \right) g^{\mu\nu} \Big] .
\end{eqnarray}
The expression (\ref{4.9}) contains a mass term for the vector potential
$A_{\mu}$.  This is particularly clear in the case $\psi = \phi$ and
$B_{\mu}=0$ for which (\ref{4.9}) reads
\begin{equation}\label{4.9a}
  \int_{M^{4}} \rmd^{4} x \sqrt{-g}
  e^{\phi} \left[ \Lambda + \Sigma + \Sigma \left(A_{\mu} +
  \partial_{\mu} \sigma \right)
    \left( A_{\nu} + \partial_{\nu} \sigma \right) g^{\mu\nu} \right] .
\end{equation}
The functional (\ref{4.9a}) consists of two parts, a 4d cosmological term with
the cosmological constant $\Lambda +\Sigma$ and a mass term for the vector
potential in the Stueckelberg formalism where the photon mass is given by
\begin{equation}\label{4.10}
  m = \hbar \sqrt{\Sigma} .
\end{equation}
Other choices of the fields $B_{\mu}$ and $\psi$ are possible.  $B_{\mu}$ may
be interpreted as an external current.

In summary, we see that from the viewpoint of a 5d unification, the
cosmological constant and the photon mass are of the same origin.

\section{Conclusions}

The main achievement of Kaluza-Klein theory is the unification of 4d
gravitational concepts and notions from Maxwell theory into 5d gravitational
concepts. This has so far been done, e.g., for the 4d metric and the vector
potential which are unified in the 5d metric, and for the momentum of a point
particle and its charge which are unified in a 5d momentum. In this article,
this Kaluza-Klein programme has been extended in two ways. First, the 4d
translational gauge symmetry of gravitation and the $U(1)$ gauge symmetry of
Maxwell theory are unified in a 5d translational symmetry. Secondly, it has
been shown that the cosmological constant and the photon mass can be unified
in a 5d cosmological tensor.

Whether the latter unification is of a deeper significance, cannot be shown by
the classical methods used in this paper. However, a further test of the
usefulness of the approach would be the generalization to nonabelian gauge
theories. Heuristically, if we consider the gauge theory of electroweak
interactions, the minimal choice for the internal manifold in a higher
dimensional unification is the homogeneous space $U(1)\times SU(2)/U(1)$.
Since this space is three-dimensional, the incorporation of a cosmological
term along the lines of section 4 of this article would lead to three mass
terms for the four gauge bosons, which qualitatively agrees with the mass
spectrum of the gauge bosons of electroweak interactions.


\begin{thebibliography}{99}
  
\bibitem{kaluza}Kaluza T 1921 {\it Sitzungsber.\ Preuss.\ Akad.\ Wiss.\ 
    Berlin} 966
  
\bibitem{klein}Klein O 1926 {\it Z.\ Phys.} {\bf 37} 895 
    
\bibitem{kohler1}Kohler C 2000 {\it Int.\ J.\ Mod.\ Phys.} {\bf A 15} 1235
  
\bibitem{kobayashi}Kobayashi S and Nomizu K 1963 {\it Foundations of
    Differential Geometry} vol 1 (Wiley, New York)

\bibitem{kohler2}Kohler C 2000 {\it Gen.\ Rel.\ Grav.} {\bf 32} 1301

\bibitem{trautman}Trautman A 1973 {\it Symp.\ Math.} {\bf 12} 139
  
\bibitem{grignani}Grignani G and Nardelli G 1992 {\it Phys.\ Rev.} {\bf D 45}
  2719

\bibitem{stueckelberg}Stueckelberg E C G 1938 {\it Helv.\ Phys.\ Acta} {\bf
    11} 299
  
\end{thebibliography}
\end{document}